\documentclass[nofootinbib,superscriptaddress]{revtex4}

\usepackage{amssymb}
\usepackage{amsmath}
\usepackage{graphicx}
\usepackage{longtable}
\usepackage{verbatim}
\usepackage{amsfonts}
\usepackage[bookmarksdepth=2]{hyperref}
\usepackage{epstopdf}
\usepackage{color}
\usepackage{units}

\newcommand{\be}{\begin{eqnarray}}
\newcommand{\ee}{\end{eqnarray}}
\newcommand{\bea}{\begin{eqnarray}}
\newcommand{\eea}{\end{eqnarray}}
\newcommand{\beq}{\begin{eqnarray}}
\newcommand{\eeq}{\end{eqnarray}}
\def\beqa{\begin{eqnarray}}
\def\eeqa{\end{eqnarray}}

\def\lsim{\mathrel{\rlap{\lower4pt\hbox{\hskip1pt$\sim$}}
    \raise1pt\hbox{$<$}}}         
\def\gsim{\mathrel{\rlap{\lower4pt\hbox{\hskip1pt$\sim$}}
    \raise1pt\hbox{$>$}}}         

\def\R{\mathcal{R}}

\def\X{X_{\rm esc}}

\begin{document}

\vspace*{-10mm}

\title{Rapidity dependence of nuclear coalescence: impact on cosmic ray antinuclei
}

\author{Kfir Blum}\email{kfir.blum@weizmann.ac.il}
\affiliation{Department of Particle Physics \& Astrophysics, Weizmann Institute of Science, Rehovot, Israel}

\vspace*{1cm}

\begin{abstract}
Upcoming studies at the Large Hadron Collider (LHC) aim to extend the rapidity coverage in  measurements of the production cross section of antinuclei ${\rm \bar d}$ and $\overline{^3\rm He}$. We illustrate the impact of such studies on cosmic ray (CR) flux predictions, important, in turn, for the interpretation of results from CR experiments. We show that, in terms of the rapidity effect, covering the range $|y|<1.5$ at the LHC should be sufficient for the astrophysical CR calculation. Important extrapolation remains in other aspects of the problem, notably $\sqrt{s}$. 
\end{abstract}

\maketitle

\section{Introduction and result}
Secondary astrophysical antimatter is produced in the galaxy in a fixed-target set-up with high-energy primary cosmic rays (CRs, mostly p) scattering on ambient gas (mostly p). Detecting this astrophysical flux (already achieved for ${\rm \bar p}$ and ${\rm e^+}$) is one of the main goals of CR experiments~\cite{Simon:2012pta,Berdugo:2022zzo}. The production cross sections for antimatter particles formed in pp collisions are not calculable from first principles, and must be obtained from accelerator experiments. 
A general obstacle is the limited kinematical coverage of these experiments (see, e.g.~\cite{Blum:2017qnn,Blum:2017iol,Korsmeier:2018gcy}). This becomes particularly important for antinuclei like $\bar{\rm d}$ and $\overline{^3\rm He}$. Ref.~\cite{Blum:2017qnn} demonstrated the impact of $p_t$ and $\sqrt{s}$ coverage for $\bar{\rm p}$, $\bar{\rm d}$, and $\overline{^3\rm He}$ production (see App.~A {\it there}), but the rapidity dependence was not exhibited. Ref.~\cite{Korsmeier:2018gcy} discussed ${\rm \bar p}$ production kinematics, including rapidity (or Feynman $x$), but did not consider $\bar{\rm d}$ and $\overline{^3\rm He}$. 
This paper illustrates the impact of rapidity coverage on the production of these anti-nuclei, in anticipation of experimental progress~\cite{Citron:2018lsq}.  

\begin{figure}[!h]\begin{center}
\includegraphics[width=0.495\textwidth]{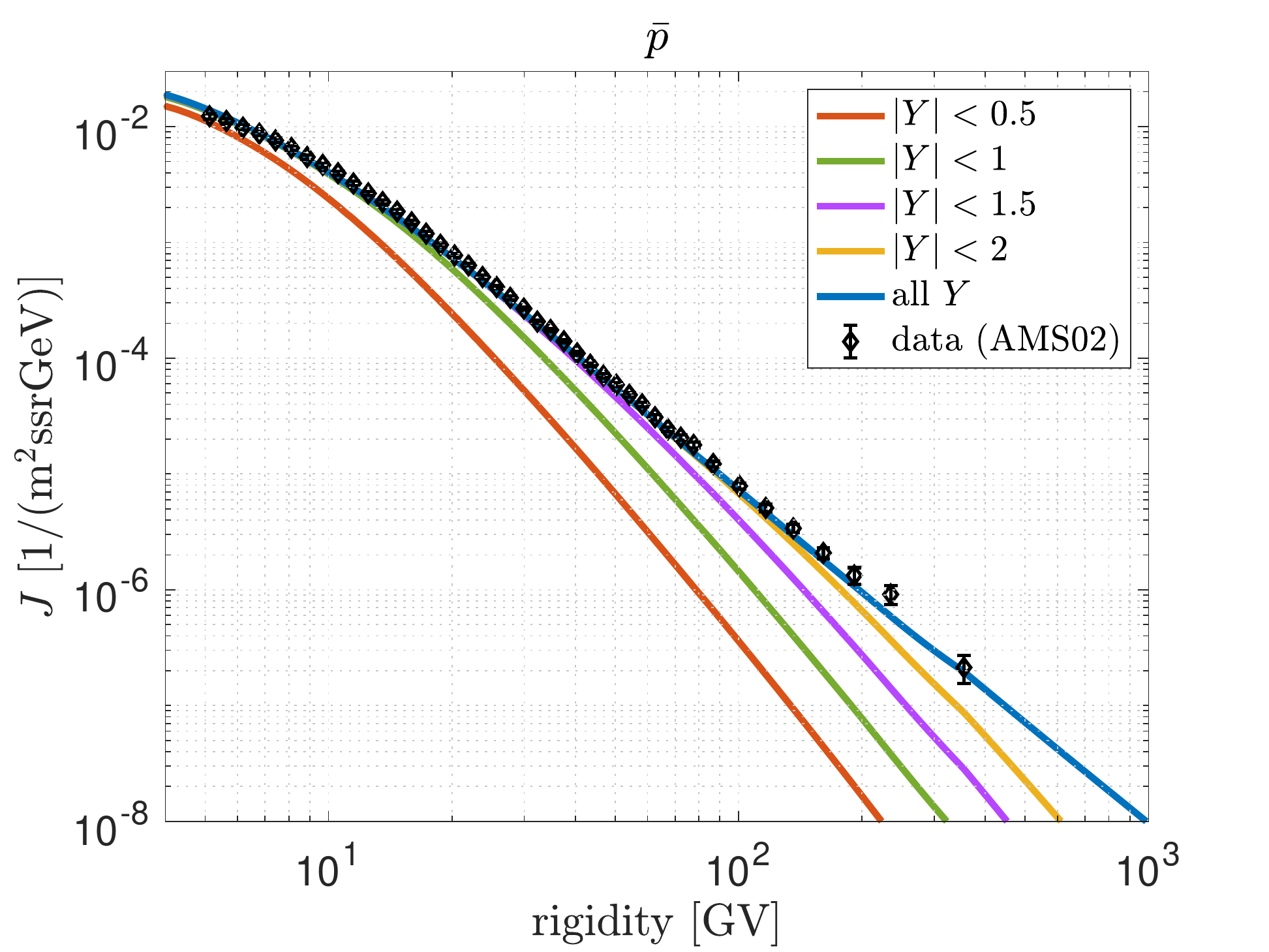}
\includegraphics[width=0.495\textwidth]{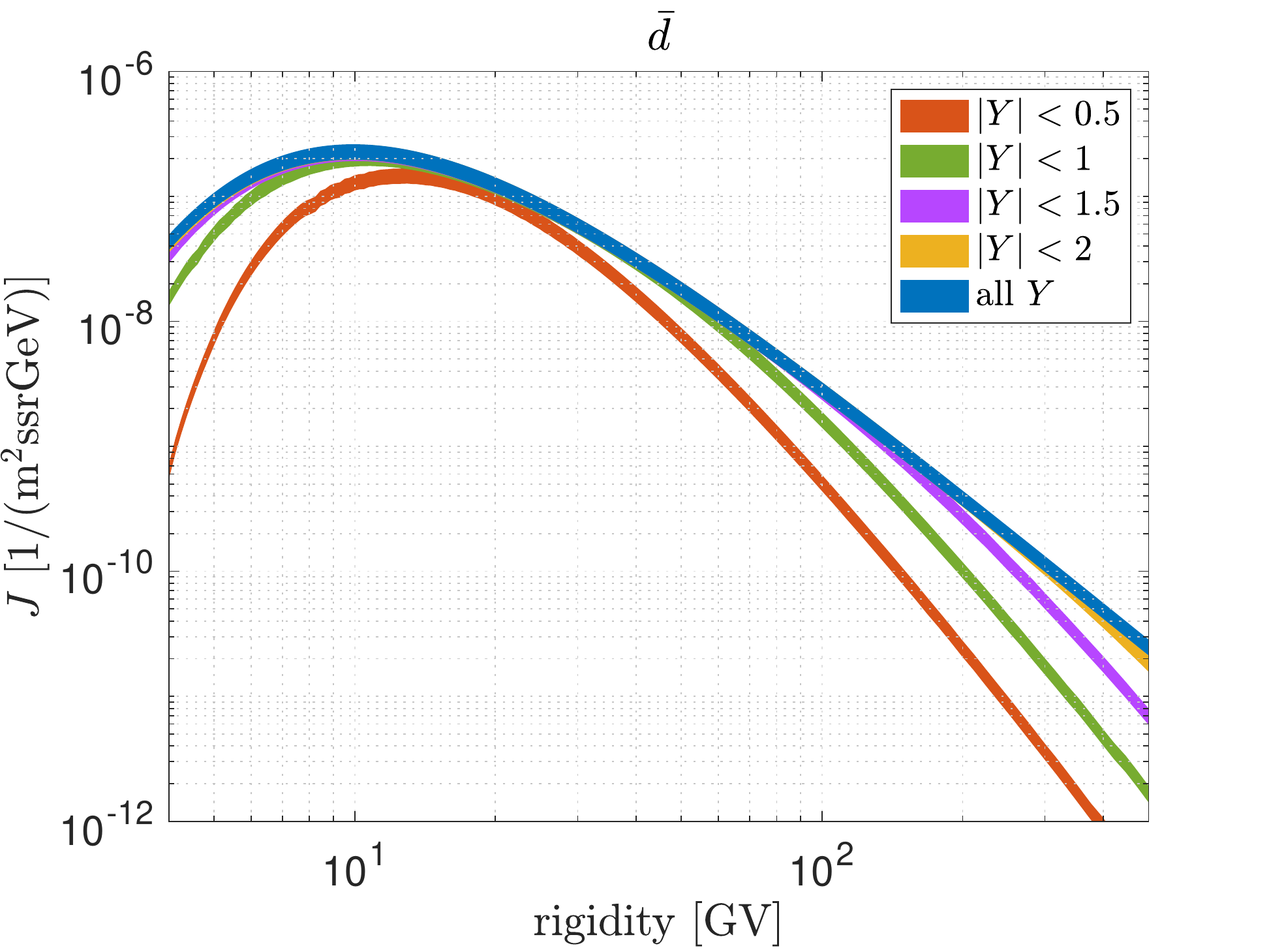}
\includegraphics[width=0.495\textwidth]{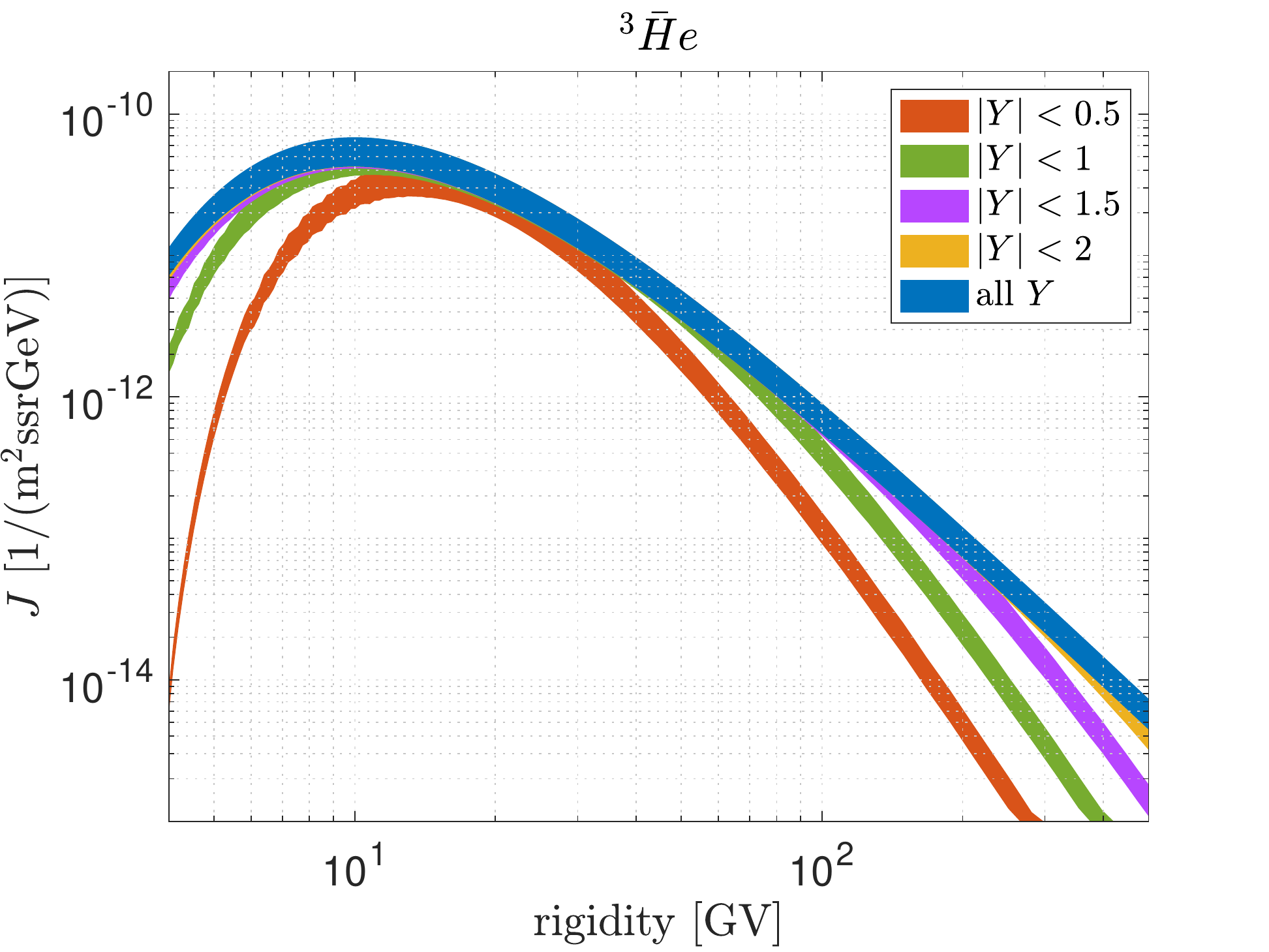}
\end{center}
\caption{Impact of rapidity cut on predicted CR antinuclei fluxes.
}\label{fig:rapcut}
\end{figure}

At the time of writing, accelerator yields of ${\rm \bar d}$ and $\overline{^3\rm He}$ are only reported for $|y|<0.5$~\cite{ALICE:2019dgz,ALICE:2020foi,Acharya:2017fvb,ALICE:2021ovi,ALICE:2021mfm}.  
However, the prediction of the CR flux arising from hadronic collisions is obtained by integration over all kinematical configurations leading to a given CR energy, including kinematical regions that fall outside of the measured range in rapidity. The theoretical prediction is therefore based to some extent on extrapolation. 
A simple way to assess the extent of the extrapolation is by setting the differential production cross section to zero in the relevant kinematical region, and inspecting the impact on the flux. 
We review the calculation in the body of this paper; here, to first motivate our analysis, we fast-forward to the main result, summarized by Fig.~\ref{fig:rapcut}. 

In each panel of Fig.~\ref{fig:rapcut}, dealing with $\bar{\rm p}$, ${\rm \bar d}$ and $\overline{^3\rm He}$, different curves show the CR flux that is obtained when in the kinematical integration of the CR yield we set the differential cross section to zero outside of a certain range in rapidity, noted in the legend. 
The width of the bands in the plot reflects the experimental uncertainty on the coalescence factors~\cite{Bellini:2020cbj} from~\cite{ALICE:2019dgz,ALICE:2020foi,ALICE:2021ovi,ALICE:2021mfm}, and does not include other uncertainties, of which the dominant ones are due to the nucleonic ($\bar{\rm p}$) cross section (of the order of $10-20\%$~\cite{Blum:2017qnn,Blum:2017iol}) and astrophysical propagation in the galaxy (again roughly $10-20\%$, provided that we restrict the calculation to relativistic CRs with magnetic rigidity $\R=p/eZ$ of few GV and above, and assuming that the main propagation effect is calibrated using measurements of heavier secondary CRs like the boron/carbon flux ratio~\cite{Blum:2017iwq}). 

We emphasize two lessons from Fig.~\ref{fig:rapcut}, focusing on ${\rm \bar d}$ and $\overline{^3\rm He}$:
\begin{enumerate}
\item At rigidity $\R\lesssim5$~GV, about 90\% of the CR ${\rm \bar d}$ and $\overline{^3\rm He}$ flux predictions hinges on extrapolation.\\ 
The effect is less substantial near $\R\approx10$~GV, and becomes prominent again at $\R\gtrsim100$~GV.
\item LHC experiments could essentially close the rapidity gap if they provide antinuclei yields covering the range $|y|<1.5$. (Note that the astrophysical CR ${\rm \bar d}$ flux at $\R>100$~GV may be very challenging to measure, regardless of cross section details.)
\end{enumerate}    

Even with rapidity effects settled, we stress that LHC experiments in collider mode\footnote{But see the fixed-target programs of~\cite{AFTERLHCstudy:2018soc,Hadjidakis:2019vpg} and, notably, the low $\sqrt{s}$ experiments AMBER~\cite{Seitz:2023rqm} and NA61/SHINE~\cite{Marcinek:2022wkm}.} are mostly limited to very high $\sqrt{s}\sim$~few TeV, while the bulk of CR ${\rm \bar d}$ are the product of pp collisions at $\sqrt{s}<30$~GeV. For example, the analysis in App.~A in~\cite{Blum:2017qnn} (see Fig.~7 {\it there}) shows that for CR at $\R=10$~GV only about $6\%,\,15\%$, and $30\%$ of $\bar{\rm p},\,\bar{\rm d}$, and $\overline{^3\rm He}$, respectively, come from initial states with $\sqrt{s}>25$~GeV. In this important sense, the CR calculation is rooted in extrapolation, regardless of rapidity. Nevertheless, it would obviously be reassuring to see at least the high $\sqrt{s}$ region of the production cross section brought under control.
 
Sec.~\ref{s:outl} contains an outline of our calculation. Sec.~\ref{ss:rapc} explains the rapidity cut. Sec.~\ref{ss:prop} explains the treatment of CR propagation (based on~\cite{Blum:2017qnn} and references in it). We summarize in Sec.~\ref{s:sum}.

\section{Outline of the calculation}\label{s:outl}
The production rate density of antinucleus species $a$ ($a=\bar{\rm d},\,\overline{^3\rm He}$) in the galaxy per unit column density of interstellar matter (ISM) target, via pp collisions, is given by~\cite{Blum:2017qnn}\footnote{For $\overline{^3\rm He}$, one needs to add the contribution of $\bar{\rm t}$ which decays in flight, and is produced with a similar cross section~\cite{ALICE:2021mfm}.}
\be\label{eq:Q} Q_a^{\rm pp}(\R)&=&\frac{1}{m}\int_{\epsilon_a}^\infty d\epsilon_{\rm p}n_{\rm p}(\epsilon_{\rm p})\frac{d\sigma_{{\rm pp}\to a}(\epsilon_{\rm p},\epsilon_a)}{d\epsilon_a}.\ee
Up to small corrections from nuclear binding energies, the final state rigidity $\R$ (the convenient quantity when dealing with  CR propagation) is related to the final state energy via $Z^2\R^2=\epsilon_a^2-A^2m_{\rm p}^2$, where $A,Z$ are the mass and charge numbers. The integral is performed over $\epsilon_{\rm p}$, the observer frame energy of the incoming primary proton, with primary protons number density $n_{\rm p}(\epsilon_{\rm p})$. The mean ISM target nucleon mass is $m\approx1.3m_{\rm p}$. The observer frame energy of the product is $\epsilon_a$. 
The key object is the Lorentz-invariant differential cross section $\left(\epsilon_a\frac{d\sigma_{{\rm pp}\to a}}{d^3{\bf p}_a}\right)$, which enters as
\be\label{eq:cs}\frac{d\sigma_{{\rm pp}\to a}(\epsilon_{\rm p},\epsilon_a)}{d\epsilon_a}&=&2\pi p_a\int dc_\theta\left(\epsilon_a\frac{d\sigma_{{\rm pp}\to a}}{d^3{\bf p}_a}\right).\ee
Here, $p_a=|{\bf p}_a|$ is the magnitude of the observer frame 3-momentum of $a$, and $\theta$ is the observer frame scattering angle. The transverse momentum is 
$p_{at}=p_as_\theta$. 
We abbreviate $c_\theta=\cos\theta,\,s_\theta=\sin\theta$.

As far as astrophysics is concerned, it would be desirable to receive direct accelerator  measurements of $\left(\epsilon_a\frac{d\sigma_{{\rm pp}\to a}}{d^3{\bf p}_a}\right)$ in as many kinematic configurations as possible. 
In practice one attempts to reduce the required kinematics coverage by using a coalescence prescription~\cite{Scheibl:1998tk,Blum:2019suo,Bellini:2020cbj},
\be \left(\epsilon_a\frac{d\sigma_{{\rm pp}\to a}}{d^3{\bf p}_a}\right)&=&R_A(\tilde\epsilon_X)B_A({\bf p}_{\rm p})\left(\epsilon_{\rm \bar p}\frac{d\sigma_{{\rm pp}\to \bar p}}{d^3{\bf p}_{\rm \bar p}}\right)^A.\ee
Here, $B_A({\bf p}_{\rm p})$ is the coalescence factor, and the prescription is useful if $B_A({\bf p}_{\rm p})$ is a slow-varying function of the kinematics. $R_A(\tilde\epsilon_X)$ is a phase-space correction factor (see App.~C of~\cite{Blum:2017qnn}). $\tilde\epsilon_X$ is the energy of the residual state produced in association to $a$ in the reaction ${\rm pp}\to aX$: $\tilde\epsilon_X=\sqrt{s+m_a^2-2\sqrt{s}\tilde\epsilon_a}$, 
defined in the pp centre of mass (CM) frame, 
where $\tilde\epsilon_a$ is the energy of $a$ in the CM frame and 
\be s&=&2m_p(m_p+\epsilon_{\rm p})\ee
is the pp collision centre of mass energy. $\left(\epsilon_{\rm \bar p}\frac{d\sigma_{{\rm pp}\to \bar p}}{d^3{\bf p}_{\rm \bar p}}\right)$ is the Lorentz-invariant production cross section for antiprotons, for which the kinematical coverage of accelerator data is usually much better than for antinuclei. We use the parameterization from~\cite{Blum:2017iol}, calibrated on LHC and other data sets.   

In Fig.~\ref{fig:rapcut}, in the ${\rm \bar d}$ calculation, the baseline value of the coalescence factor $B_2$ is taken from ALICE pp analysis at $\sqrt{s}=7$~TeV~\cite{ALICE:2019dgz}\footnote{ALICE pp analysis at $\sqrt{s}=13$~TeV~\cite{ALICE:2020foi} finds consistent, but slightly lower $B_2$.}. For $\overline{^3\rm He}$, the baseline value of $B_3$ is taken from~\cite{ALICE:2021ovi,ALICE:2021mfm}\footnote{ALICE pp analysis at $\sqrt{s}=7$~TeV~\cite{Acharya:2017fvb} finds consistent, but slightly higher $B_3$.}. We emphasize yet again that the uncertainty bands in Fig.~\ref{fig:rapcut} only include the experimental measurement uncertainties on $B_A$.

\subsection{Impact of a rapidity cut}\label{ss:rapc}
The rapidity of $a$ is defined in the CM frame as
\be y&=&\frac{1}{2}\ln\frac{\tilde\epsilon_a+\tilde p_{az}}{\tilde\epsilon_a-\tilde p_{az}}\;=\;\frac{1}{2}\ln\frac{1+\tilde
\beta_z}{1-\tilde\beta_z},\ee
so the CM frame longitudinal boost velocity is
\be\label{eq:rapbz}\tilde\beta_z&=&\tanh y.\ee
The boost factor between the observer frame and the CM frame is
\be\label{eq:gampp}
\gamma&=&\frac{\sqrt{s}}{2m_p}\;=\;\sqrt{\frac{m_{\rm p}+\epsilon_{\rm p}}{2m_{\rm p}}}.
\ee
The observer frame boost factor to the rest frame of $a$ is
\be\gamma_a&=&\frac{\epsilon_a}{m_a}\;=\;\frac{1}{\sqrt{1-\beta_a^2}},\ee
with which $p_a=\beta_a\epsilon_a=\gamma_a\beta_am_a=m_a\sqrt{\gamma_a^2-1}$.
The CM frame energy of $a$, $\tilde\epsilon_a$, is 
\be\tilde\epsilon_a&=&m_a\left(\gamma\gamma_a-c_\theta\sqrt{\gamma^2-1}\sqrt{\gamma_a^2-1}\right).\ee
Using $\tilde p_{az}^2=\tilde\epsilon_a^2-p_a^2s_\theta^2-m_a^2$, we have
\be \tilde\beta_z&=&\sqrt{1-\frac{p_a^2s_\theta^2+m_a^2}{\tilde\epsilon_a^2}}\;=\;\sqrt{1-\frac{1+(\gamma_a^2-1)s_\theta^2}{\left(\gamma\gamma_a-c_\theta\sqrt{\gamma^2-1}\sqrt{\gamma_a^2-1}\right)^2}}.\ee

An experimental cut $|y|<Y$ implies that in the CR production calculation, the regions of the integrals over $\epsilon_{\rm p},\theta$, which deviate outside of the restriction (where $\gamma=\gamma(\epsilon_{\rm p})$ via Eq.~(\ref{eq:gampp}))
\be\label{eq:Ycstrt} 1-\frac{1+(\gamma_a^2-1)s_\theta^2}{\left(\gamma\gamma_a-c_\theta\sqrt{\gamma^2-1}\sqrt{\gamma_a^2-1}\right)^2}&<&\tanh^2 Y,\ee
are based on extrapolation. 
As noted in the Introduction, we can assess to what extent  the astrophysics CR computation depends on extrapolation, by setting to zero the part of the integration which goes outside the tested kinematic region. 
This is the method we used to implement the rapidity cuts in Fig.~\ref{fig:rapcut}. 

\begin{figure}[!h]\begin{center}
\includegraphics[width=0.475\textwidth]{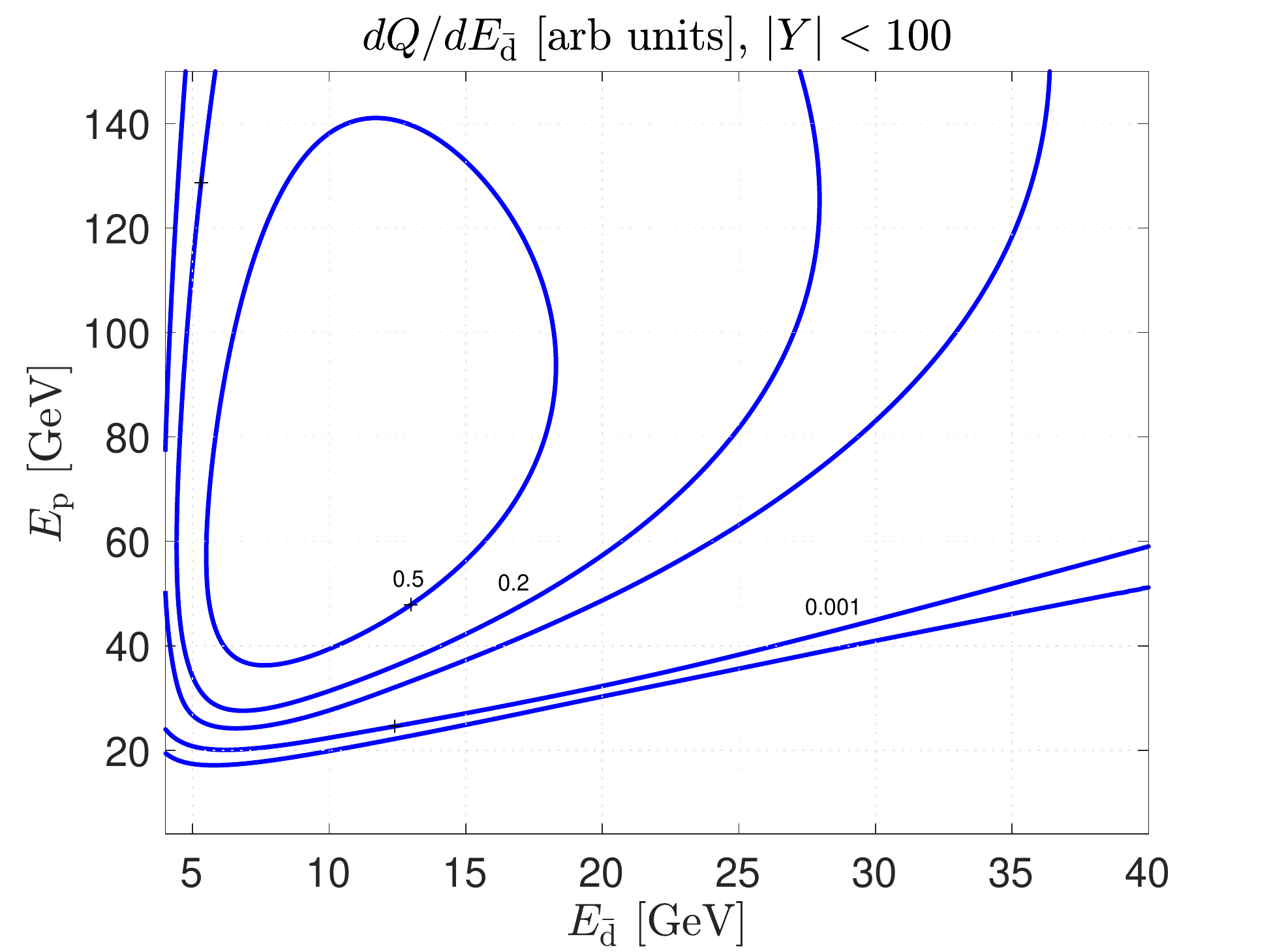}
\includegraphics[width=0.475\textwidth]{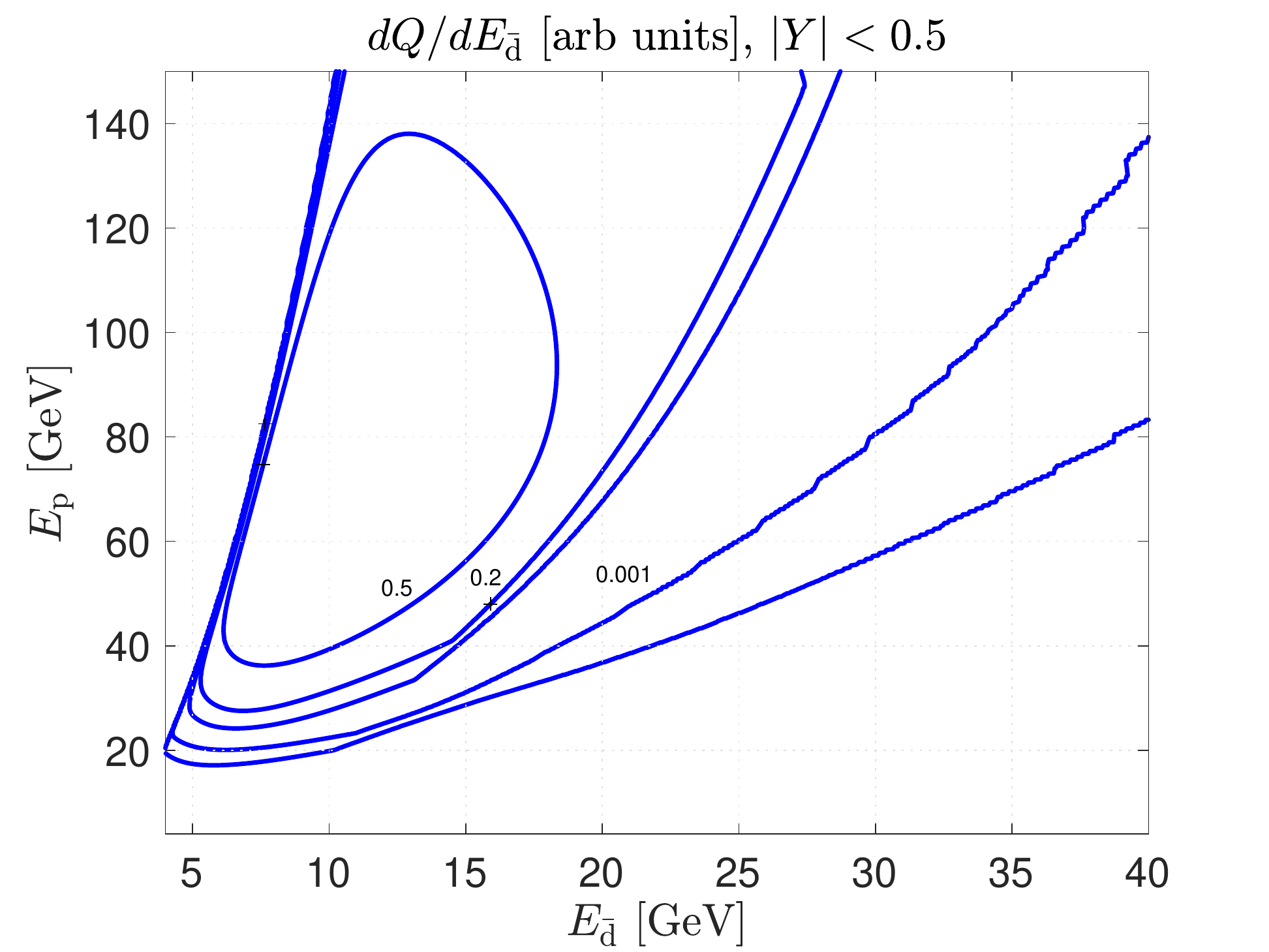}
\end{center}
\caption{Proton flux-weighted, $p_{at}$-integrated differential cross section for $\bar{\rm d}$ production. {\bf Left:} no rapidity cut. {\bf Right:} nulling kinematic regions with $|y|<0.5$.
}\label{fig:dQdep}
\end{figure}

There is a convenient approximate way to inspect the kinematics constraints in the $\epsilon_a,\epsilon_{\rm p}$ plane, glossing over the $\theta$ (or $p_{at}$) variable. 
To get there, let us first plot the integrand of Eq.~(\ref{eq:Q}), 
\be\label{eq:dEdEpa} \frac{dQ^{\rm pp}_a(\epsilon_{\rm p},\epsilon_a)}{d\epsilon_{\rm p}}=n_{\rm p}(\epsilon_{\rm p})\frac{d\sigma_{{\rm pp}\to a}(\epsilon_{\rm p},\epsilon_a)}{d\epsilon_a}.
\ee
In this definition we drop an over-all constant factor and consider the result in arbitrary units. We plot the result for $a=\bar{\rm d}$ in Fig.~\ref{fig:dQdep}. On the {\bf left} panel we show the full result without a rapidity cut. On the {\bf right} we show the result restricted to the region $|y|<0.5$. To understand the physical significance of this plot, recall that the astrophysics source term at observer frame $a$ energy $\epsilon_a$ is obtained\footnote{Up to an uninteresting propagation effect.} by integrating $\frac{dQ^{\rm pp}_a(\epsilon_{\rm p},\epsilon_a)}{d\epsilon_{\rm p}}$ over $\epsilon_{\rm p}$, the y-axis of the plot. The nulled regions in the right panel show how swaths of $\epsilon_{\rm p}$ contributing to the integral, are based on cross section extrapolation.
We see that removing high-rapidity kinematic regions suppresses production at both low and high $\epsilon_a$. 

In App.~\ref{a:kin} we show similar plots including also $\bar{\rm p}$ and $\overline{\rm ^3He}$, as well as corresponding plots that show the $p_t$-integrated differential cross section without weighing it by the proton flux (that is, $\frac{d\sigma_{{\rm pp}\to a}(\epsilon_{\rm p},\epsilon_a)}{d\epsilon_a}$).

The general shape of the differential source term is understood by considering the maximal energy available for $a$ in the CM frame:
\be\label{eq:maxga}\tilde\epsilon_a&<&\frac{4m_{\rm p}^2\gamma^2+m_a^2-M_X^2}{4m_{\rm p}\gamma},\ee
where $M_X$ is the minimal invariant mass of the residual hadronic state produced in association with $a$. (For example, if $a=\bar{\rm d}$, the minimal reaction product is ${\rm pp}\to\bar{\rm n}\bar{\rm p}\,{\rm pppn}$, so the residual hadronic state must contain at least 4 nucleons and we have $M_X\approx4m_{\rm p}$. Similarly, for $a=\overline{^3\rm He}$, we have at least ${\rm pp}\to\bar{\rm n}\bar{\rm p}\bar{\rm p}\,{\rm ppppn}$ and $M_X\approx5m_{\rm p}$.)

\begin{figure}[!h]\begin{center}
\includegraphics[width=0.4\textwidth]{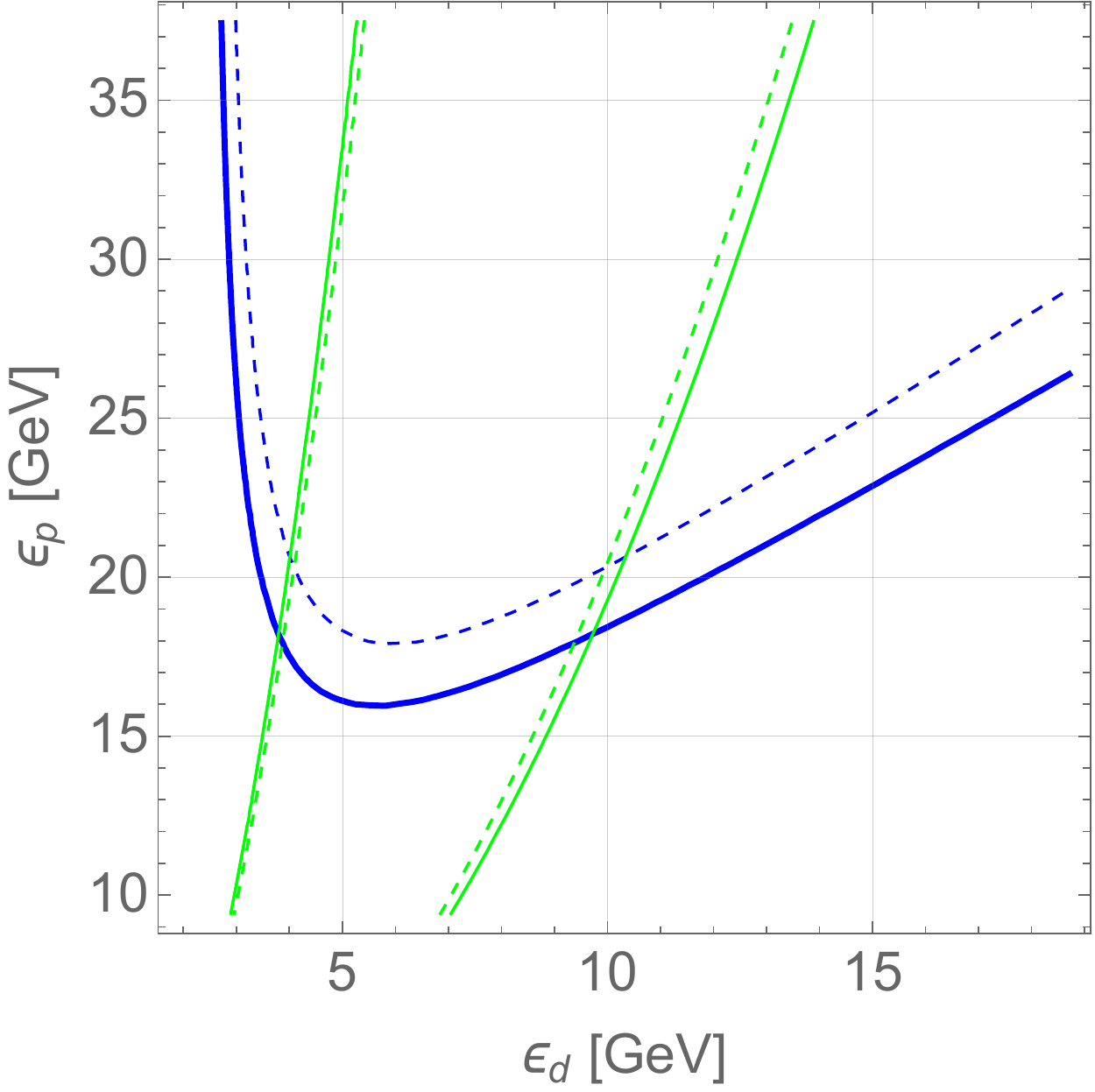}
\end{center}
\caption{Kinematics and rapidity regions for $\bar{\rm d}$ production. The generally allowed kinematic region is above the blue solid line (in the direction of the dashed line). The region $|y|<0.5$ is between the green solid lines (in the direction of the dashed line).
}\label{fig:dQde}
\end{figure}

In Fig.~\ref{fig:dQde}, above the solid blue contour Eq.~(\ref{eq:maxga}) is satisfied, and below it, it is not. This blue contour delineates the general shape of the differential source in the left panel of Fig.~\ref{fig:dQdep}: near the edge of the region, this source is shaped essentially by phase space.

The rapidity coverage can now be easily understood approximately by noting that most of the secondary production occurs not far from the forward region, $c_\theta\approx1$. Setting $c_\theta=1,\,s_\theta=0$ in Eq.~(\ref{eq:Ycstrt}), we plot the contour $\tanh^2y=\tanh^2Y$ (with $Y$ the rapidity cut) in the green line in Fig.~\ref{fig:dQde}. Between the green lines (in the direction of the dashed line), Eq.~(\ref{eq:Ycstrt}) is satisfied. This approximate consideration (glossing over the $p_t$ or $\theta$ integral) explains the shape of the right panel of Fig.~\ref{fig:dQdep}.

\subsection{Treatment of CR propagation}\label{ss:prop}
To make Fig.~\ref{fig:rapcut} we follow Ref.~\cite{Blum:2017qnn} in computing the antinuclei CR flux with and without rapidity cuts. The calculation of predicted antinuclei flux uses the measured flux of CR nuclei to calibrate the rigidity-dependent mean column density of CRs~\cite{Katz:2009yd}. A good fit to the B/C flux ratio measured by the AMS02 experiment in the range $4~{\rm GV}\lesssim\R\lesssim10^3~{\rm GV}$ is provided by
\be
\X&\approx&10.5\left(\frac{\R}{10~\rm GV}\right)^{-0.56}\left(1+\left(\frac{\R}{3\times10^3~\rm GV}\right)^{0.6}\right)~{\rm{g/cm^2}}.
\ee
The differential number density of relativistic ($\R\gg1$~GV) antinuclei is then predicted via
\be\label{eq:CRflux} n_{\rm a}(\R)&\approx&\frac{Q_{\rm a}(\R)\X(\R)}{1+\frac{\sigma_a}{m}\X(\R)},\ee
where $\sigma_a$ is the inelastic cross section for antinucleus-ISM collisions. The source term $Q_{\rm a}$ is a sum of different production channels, including proton-proton and proton-helium collisions: $Q_{\rm a}=Q^{\rm pp}_{\rm a}+Q^{\rm p\,He}_{\rm a}+Q^{\rm He\,p}_{\rm a}$. All are proportional to the same underlying pp cross section and are affected by the rapidity dependence of $B_A$ in approximately the same way.

For ${\rm \bar p}$, our prediction can be compared to measured data, shown in the {\bf top left} panel of Fig.~\ref{fig:rapcut}.

More model-dependent calculations of CR propagation are often used in the literature. These details are not essential for the impact of rapidity coverage, and we expect other astrophysical model calculations to reach similar conclusions to ours.

\section{Summary}\label{s:sum}
Progress in cosmic ray (CR) measurements in space~\cite{Berdugo:2022zzo} makes the prediction of CR antimatter fluxes very timely. This comes at fantastic coincidence with the LHC program~\cite{Citron:2018lsq}. We clarified the importance of rapidity coverage in the experimental measurements of antinuclei production cross sections ($p_t$ and $\sqrt{s}$ coverage was discussed in~\cite{Blum:2017qnn}). Our main result is that extending the range of rapidity coverage at the LHC to $|y|>0.5$ is essential for CR antinuclei predictions. Covering the range $|y|<1.5$ would settle the predictions of CR ${\rm \bar d}$ and ${\rm \overline{^3He}}$ to good accuracy. 

Even with rapidity effects settled, we stress that LHC experiments in collider mode are mostly limited to very high $\sqrt{s}\sim$~few TeV, while the bulk of CR ${\rm \bar d}$ are the product of pp collisions at $\sqrt{s}<30$~GeV (see App.~A in~\cite{Blum:2017qnn}). In this important sense, the CR calculation is rooted in extrapolation even after the rapidity effects are settled at high $\sqrt{s}$. This issue is a central goal for the fixed-target programs of~\cite{AFTERLHCstudy:2018soc,Hadjidakis:2019vpg,Seitz:2023rqm,Marcinek:2022wkm}.

\acknowledgments
I am grateful to Chiara Pinto and Alberto Caliva for pointing out the relevance of the CR production rapidity analysis for the LHC program at ALICE. This work was supported by the Israel Science Foundation grant 1784/20.

\bibliography{ref}

\begin{appendix}
\section{Additional kinematic plots}\label{a:kin}
\begin{figure}[!h]\begin{center}
\includegraphics[width=0.4\textwidth]{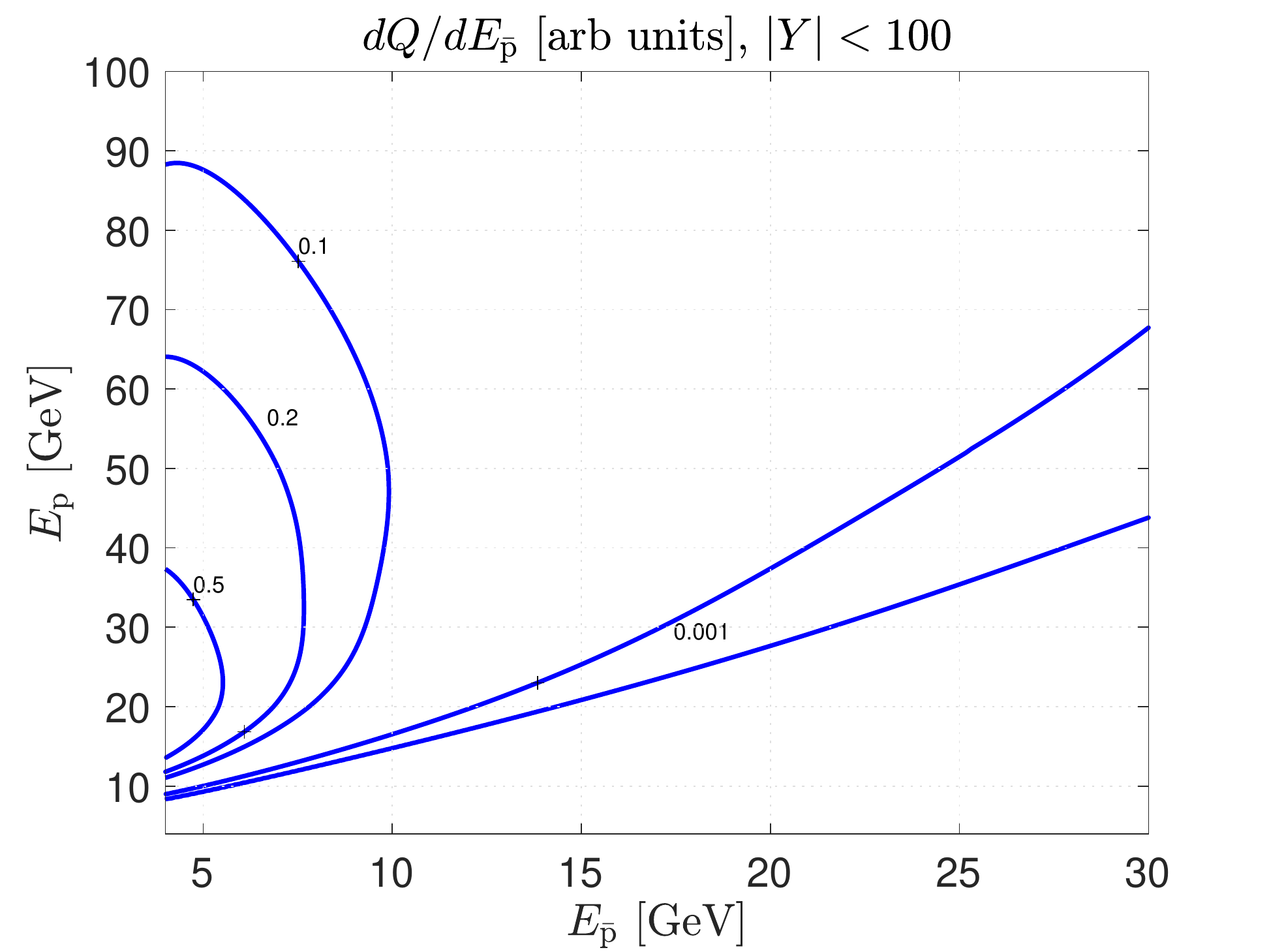}
\includegraphics[width=0.4\textwidth]{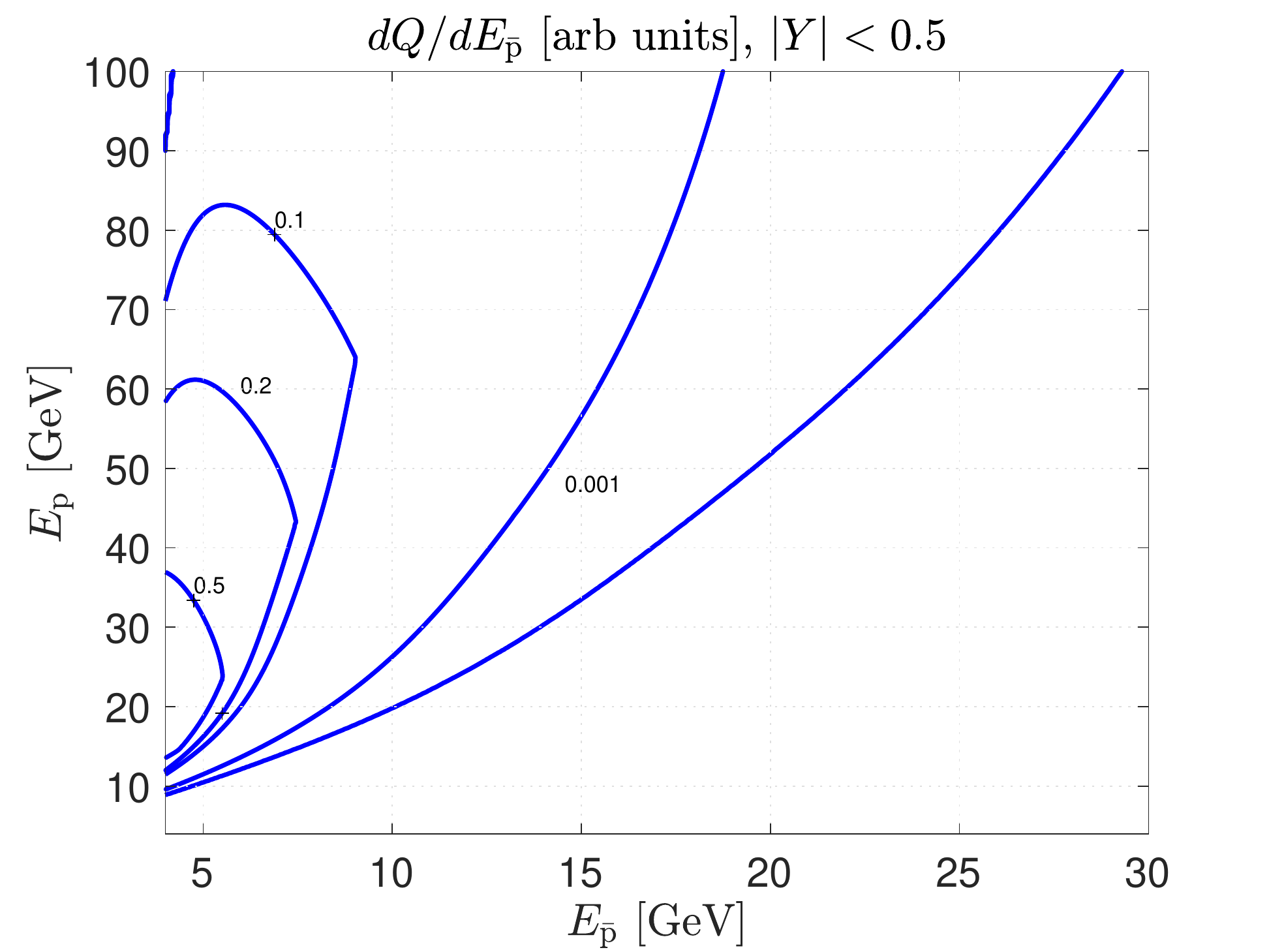}\\
\includegraphics[width=0.4\textwidth]{dQdEdbY100}
\includegraphics[width=0.4\textwidth]{dQdEdbY05}\\
\includegraphics[width=0.4\textwidth]{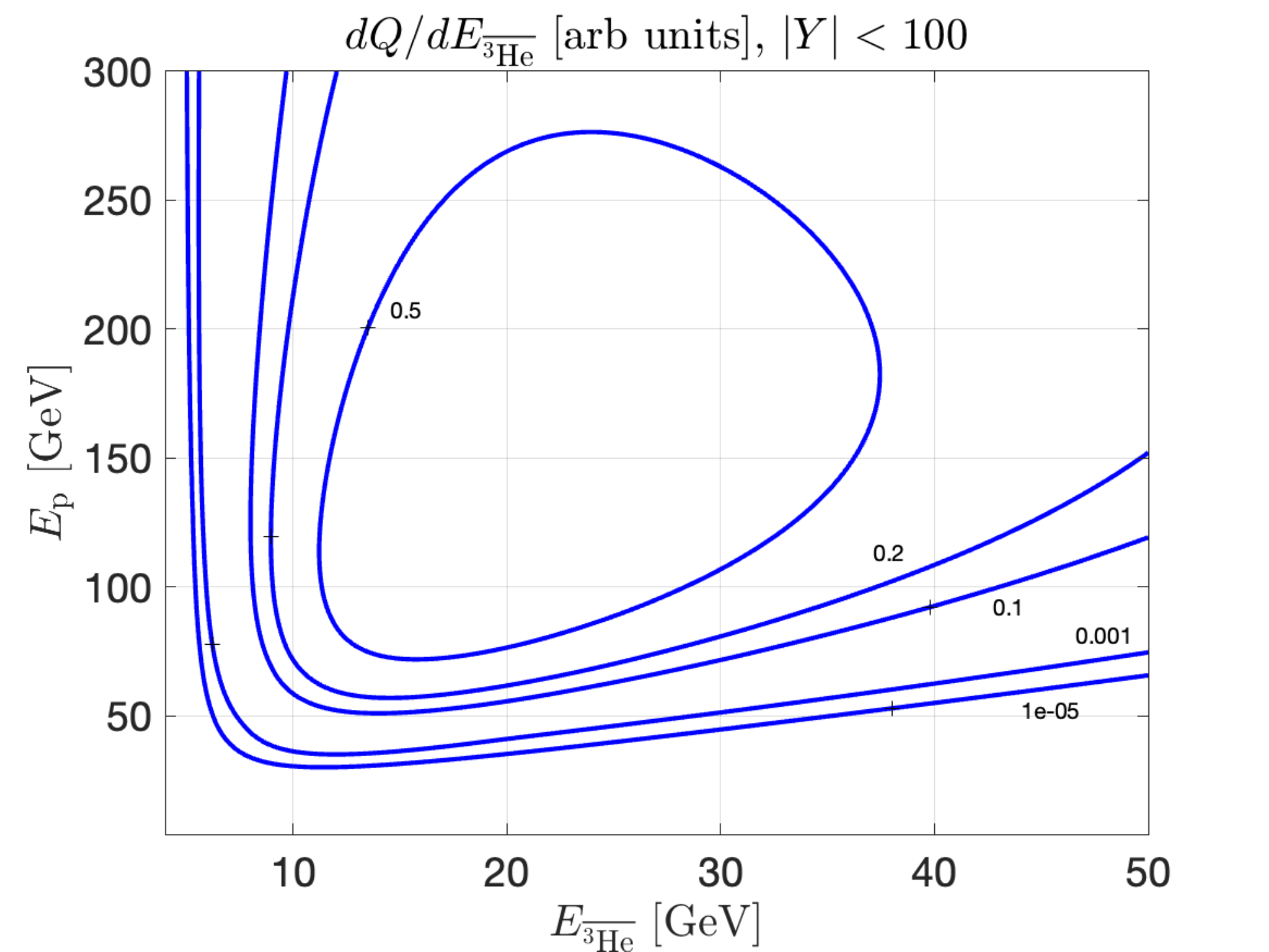}
\includegraphics[width=0.4\textwidth]{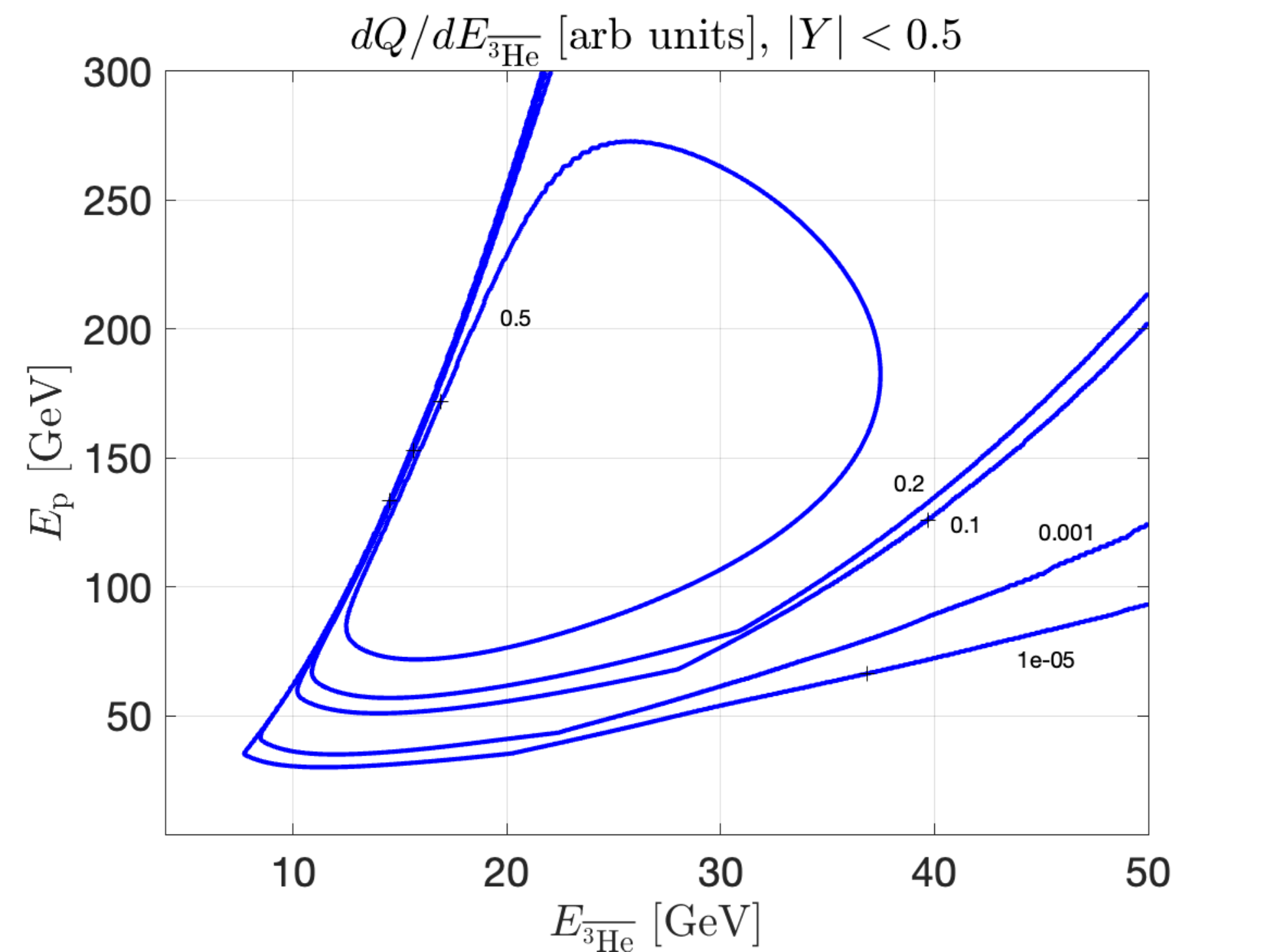}
\end{center}
\caption{Proton flux-weighted, $p_{at}$-integrated differential cross section for $\bar{\rm p}$ ({\bf top panels}), $\bar{\rm d}$ ({\bf middle panels}), and $\overline{\rm }$ ({\bf middle panels}) production. {\bf Left:} no rapidity cut. {\bf Right:} nulling kinematic regions with $|y|<0.5$.
}\label{fig:Q}
\end{figure}

\begin{figure}[!h]\begin{center}
\includegraphics[width=0.4\textwidth]{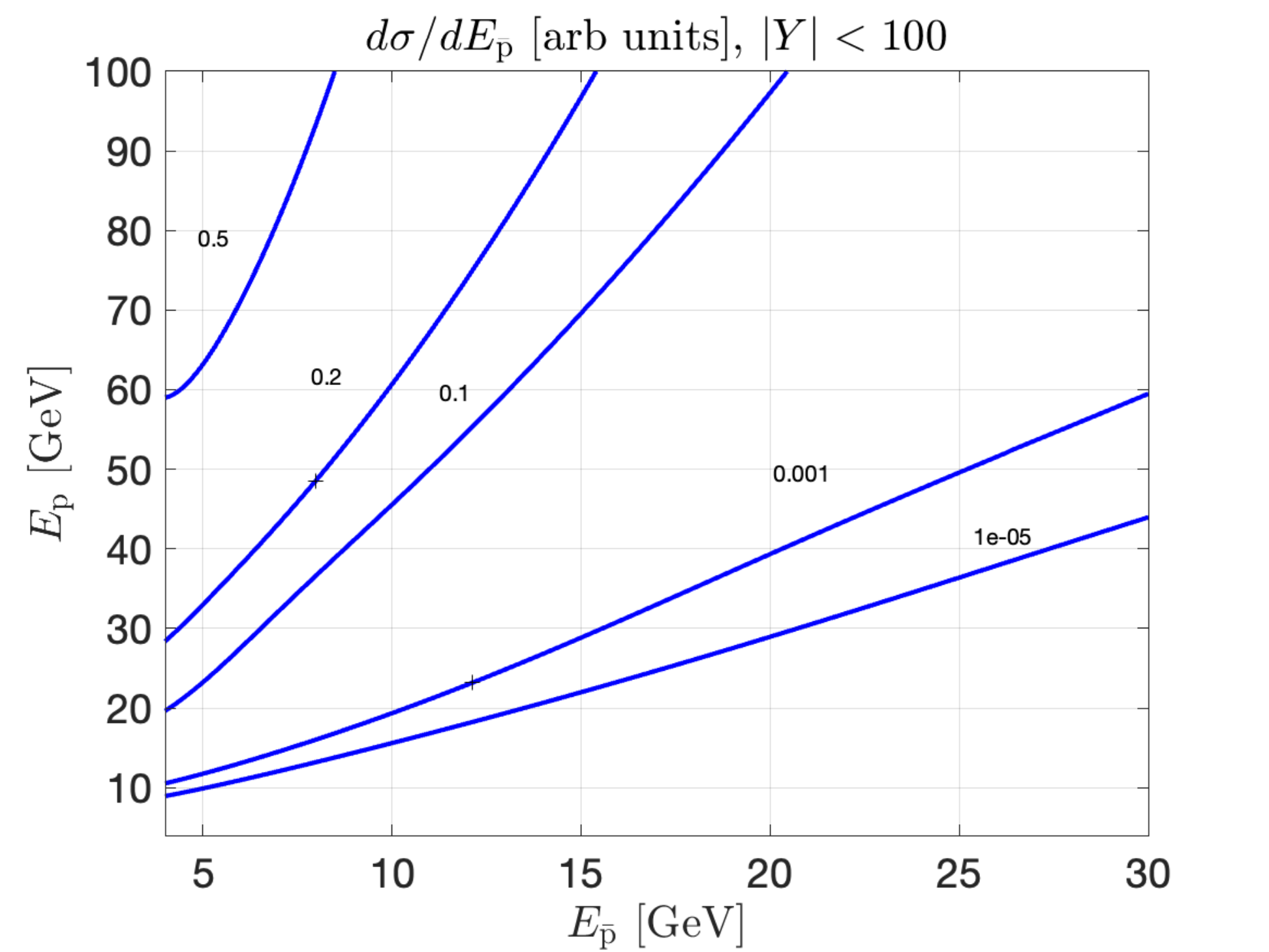}
\includegraphics[width=0.4\textwidth]{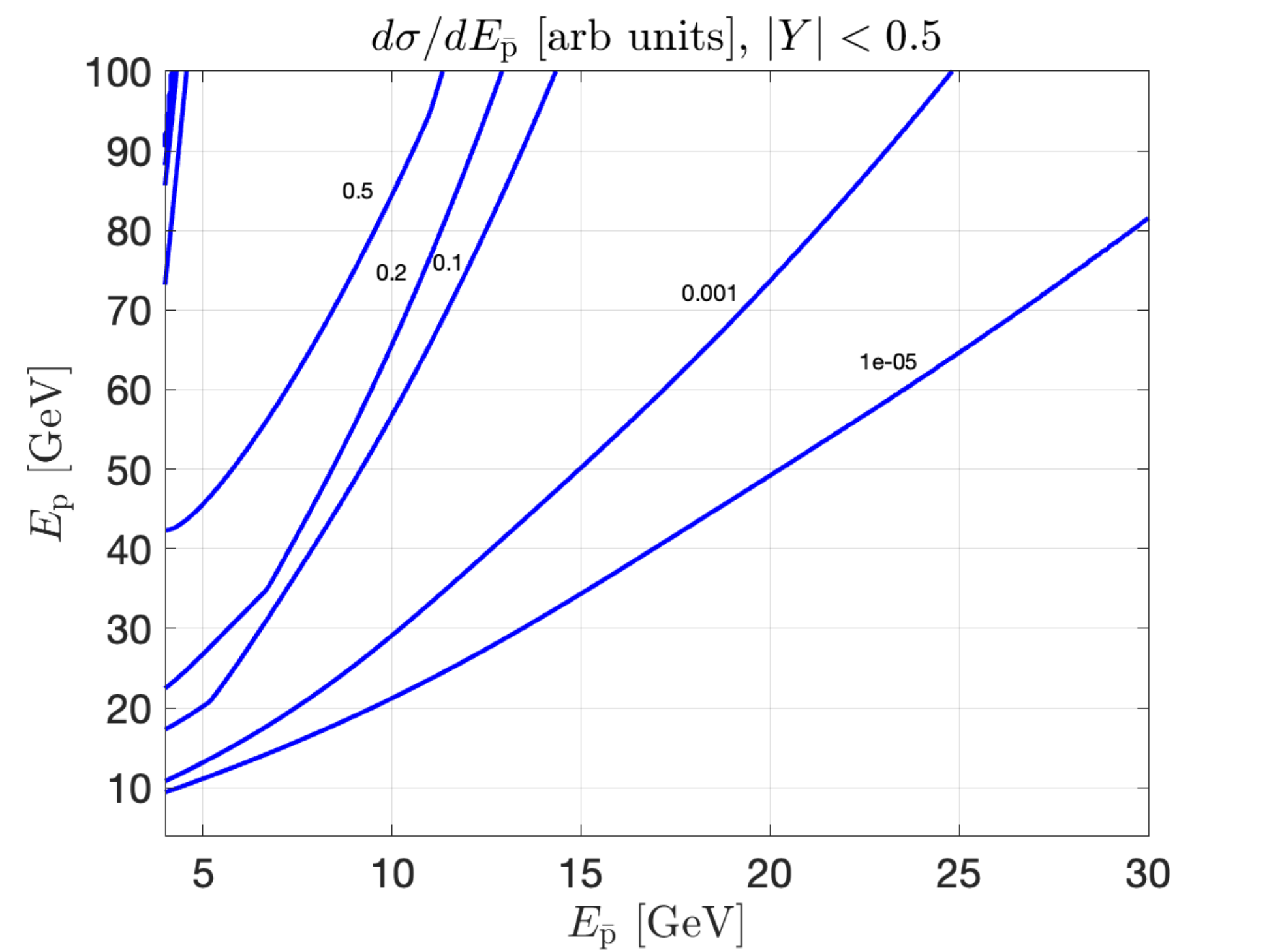}\\
\includegraphics[width=0.4\textwidth]{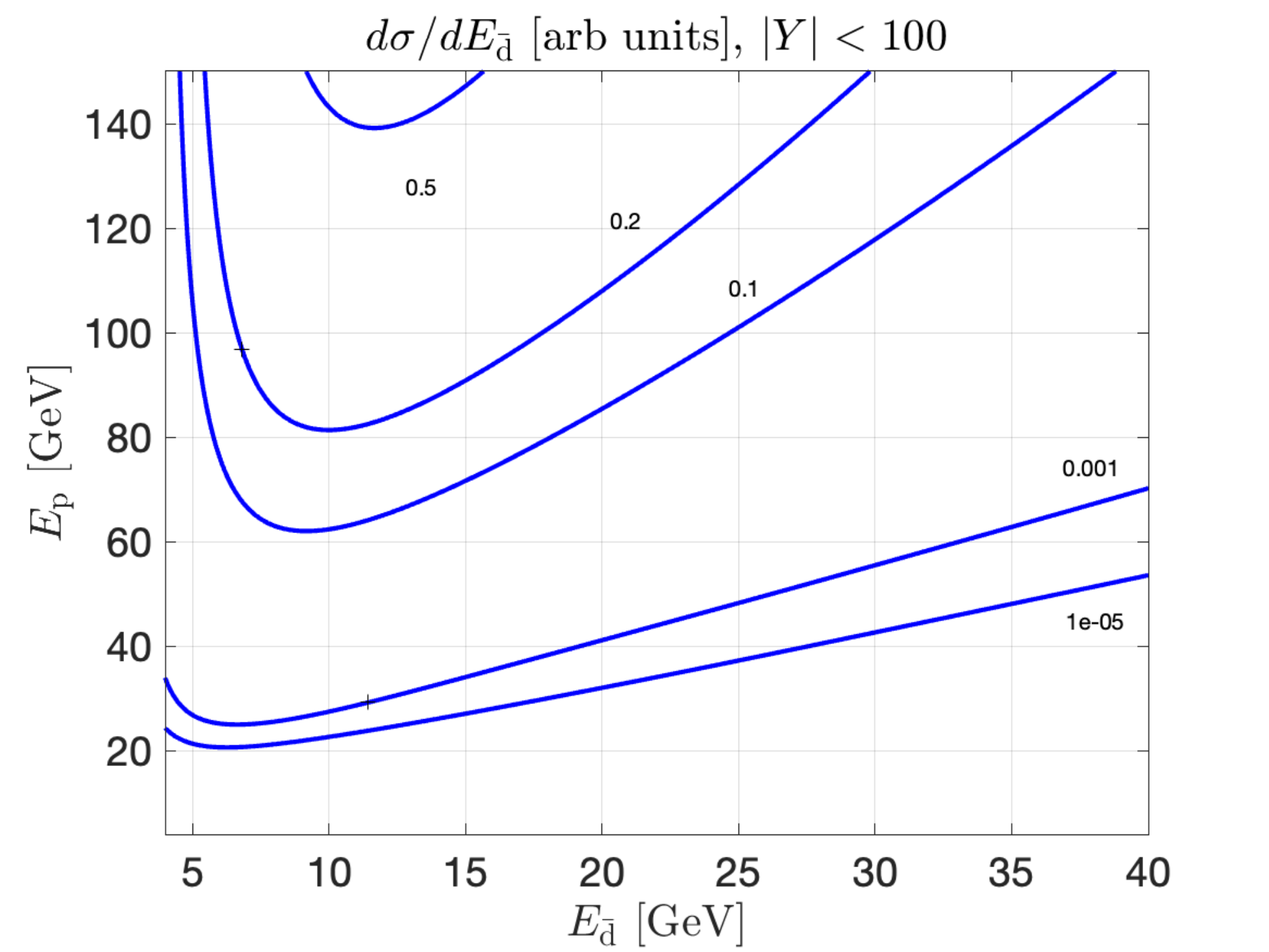}
\includegraphics[width=0.4\textwidth]{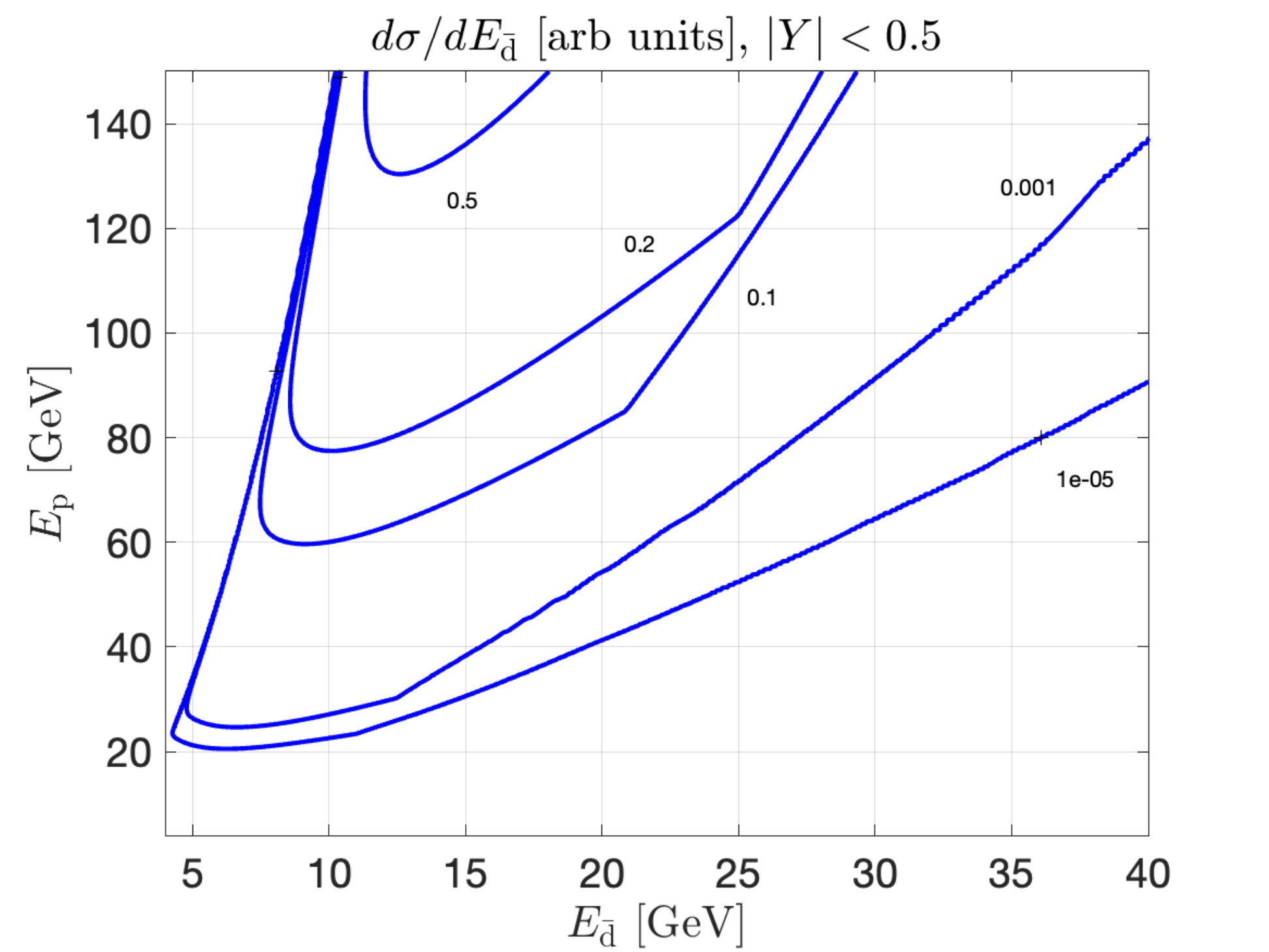}\\
\includegraphics[width=0.4\textwidth]{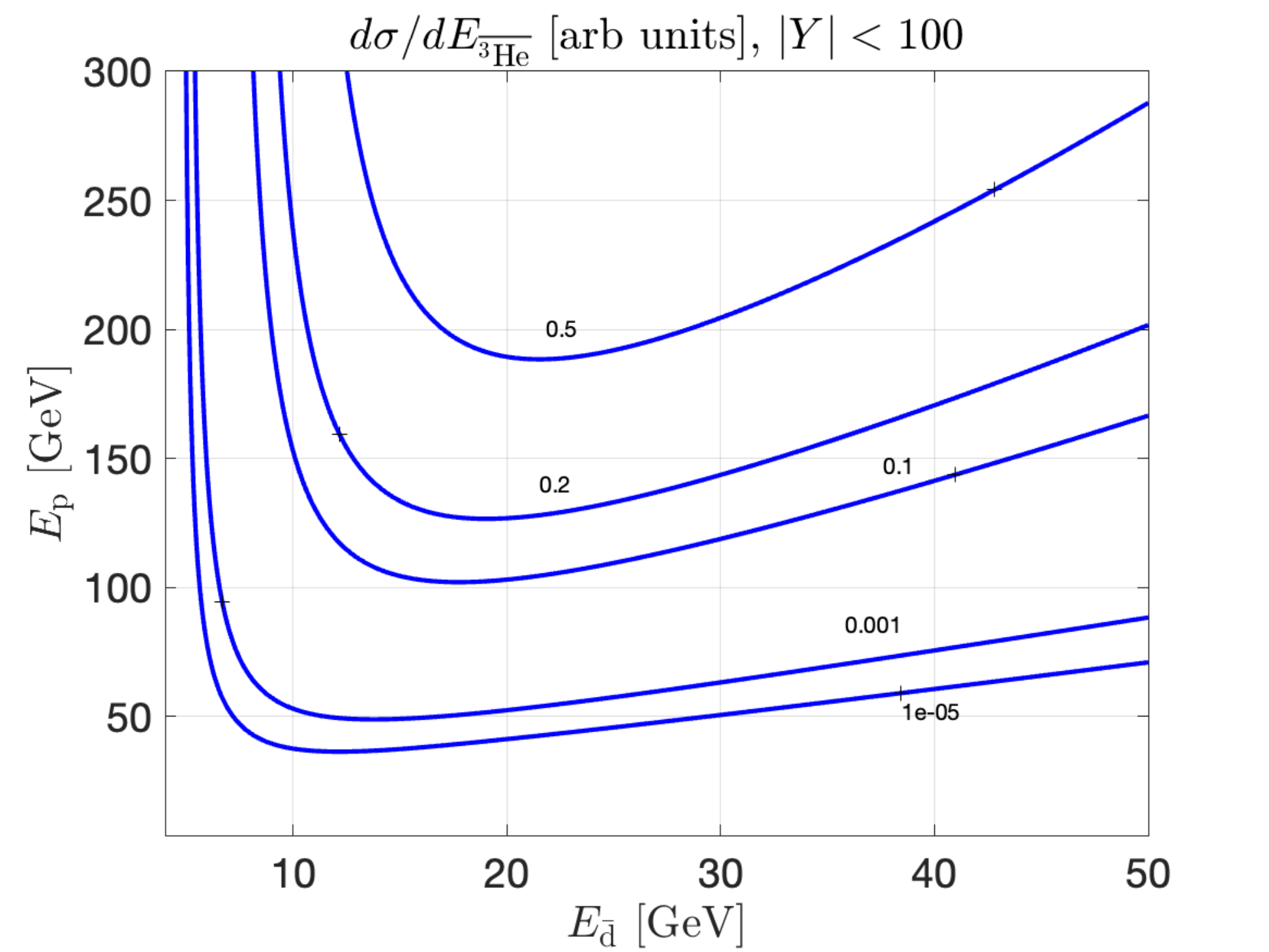}
\includegraphics[width=0.4\textwidth]{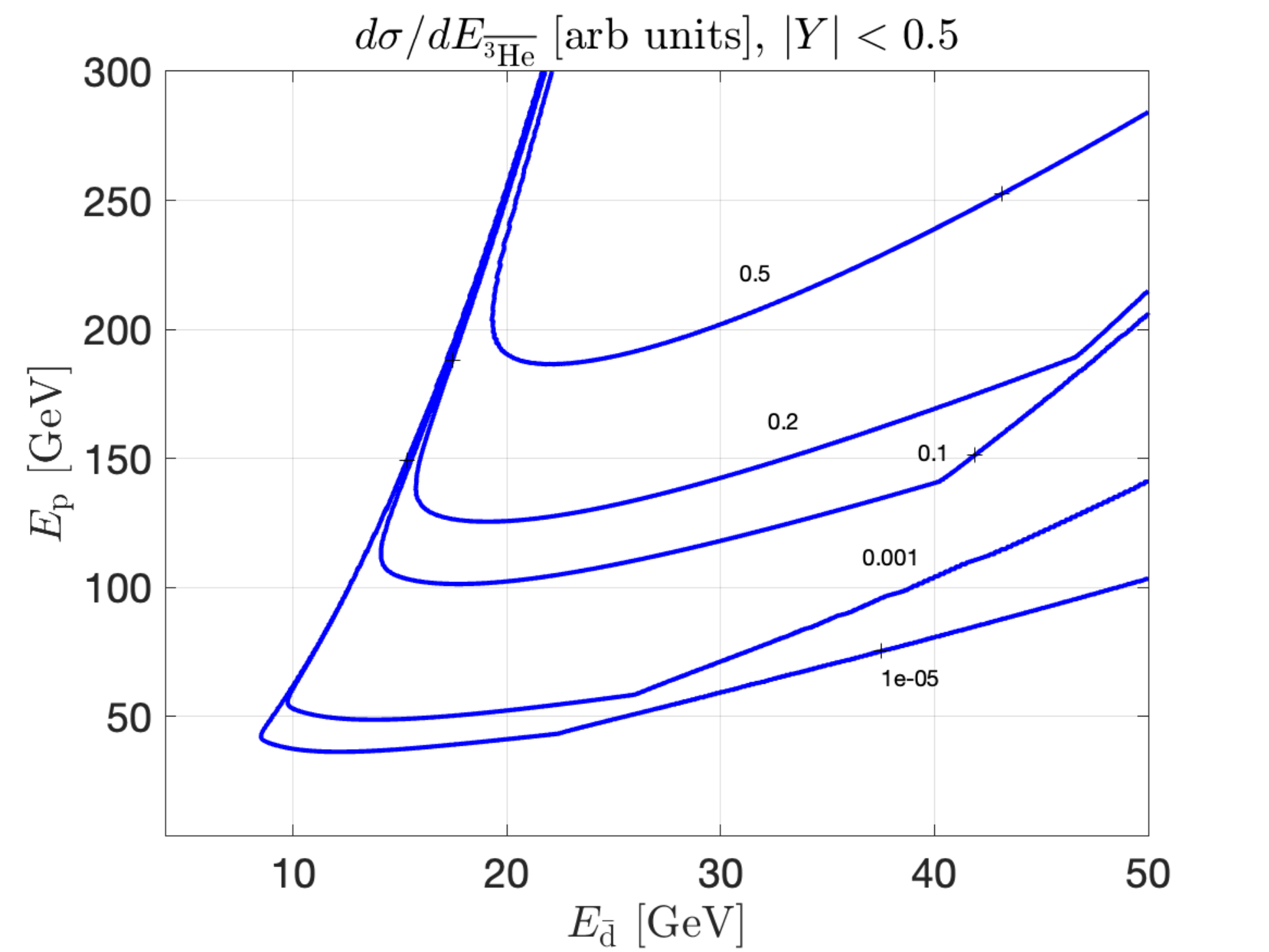}
\end{center}
\caption{$p_{at}$-integrated (NOT proton-flux weighted) differential cross section for $\bar{\rm p}$ ({\bf top panels}), $\bar{\rm d}$ ({\bf middle panels}), and $\overline{\rm }$ ({\bf middle panels}) production. {\bf Left:} no rapidity cut. {\bf Right:} nulling kinematic regions with $|y|<0.5$.
}\label{fig:s}
\end{figure}

\end{appendix}

\end{document}